# Convolutional Neural Networks for Space-time Block Coding Recognition

W. Yan, Q. Ling, L. Zhang, and Y. Zhang

We apply the latest advances in machine learning with deep neural networks to the tasks of radio modulation recognition, channel coding recognition, and spectrum monitoring. This paper first proposes an identification algorithm for space-time block coding of a signal. The feature between spatial multiplexing and Alamouti signals is extracted by adapting convolutional neural networks after preprocessing the received sequence. Unlike other algorithms, this method requires no prior information of channel coefficients and noise power, and consequently is well-suited for noncooperative contexts. Results show that the proposed algorithm performs well even at a low signal-to-noise ratio.

*Introduction:* Cognitive radio plays an important role in military applications, including electronic warfare, interference identification, and interference signal design, and has been applied in civil communications such as radio surveillance, software defined radio, and spectrum awareness [1]. Automatic signal identification is required even at a low signal-to-noise ratio (SNR) without prior information or front processing. Consequently, automatic signal identification is always the difficulty and emphasis of signal processing.

Deep neural networks (DNNs) have been widely used in computer vision, natural language processing, and speaker recognition. Researchers in the wireless communication field have started to apply DNNs to cognitive radio tasks, such as radio modulation recognition [2][3], turbo recognition, polar coding recognition [4], spectrum monitoring [5], autoencoders for end-end communication [6], and channel estimation [7]. Surprisingly, relatively simple convolutional neural networks (CNNs) outperform these traditional methods.

All space-time block coding (STBC) classification we have found in the literature has applied traditional methods such as maximum-likelihood (ML) criteria and feature-based (FB) methods [8]–[10]. These methods do not perform well in accuracy and real-time. However, the trend in machine learning applied to voice recognition and image processing is overwhelmingly that of feature learning from data rather than crafting of expert features, suggesting we should apply a similar shift in the signal-processing domain.

In this paper, we apply CNNs to STBC recognition, and we demonstrate that this approach can improve classification accuracy compared to current methods.

*STBC Signal:* Let us consider spatial multiplexing (SM) and Alamouti (AL) signals. The modulated symbols are divided by the length $N_s$, which is encoded to generate $N_t$ parallel signal sequences of length $L$. The bth block of $N_s$ complex symbols is written as a column vector, $X_b = [x_{b,0}, \cdots x_{b,N_s-1}]^T$, where the superscript T represents transposition. The coding matrix is denoted by $C(X_b)$. The transmitted symbols assumed to belong to an M-signal constellation are independent and identically distributed (IID).

For an SM signal, a block of $N_s = N_t = 2$ symbols is transmitted via the $N_t$ antennas in $L$ time periods. The coding matrix can be expressed as

$$C(X_b) = \begin{bmatrix} x_{b,0} & x_{b,2} & x_{b,4} & \cdots \\ x_{b,1} & x_{b,3} & x_{b,5} & \cdots \end{bmatrix}. \quad (1)$$

For an AL signal, a block of $N_s = 2$ is transmitted via the two antennas in $L=2$ time periods. The coding matrix can be expressed as

$$C(X_b) = \begin{bmatrix} x_{b,0} & -x_{b,1}^* \\ x_{b,1} & x_{b,0}^* \end{bmatrix}. \quad (2)$$

Without loss of generality, the first received symbols are denoted by $r(0)$, and the $(k_1+1)th$ received symbols, which belong to the bth transmitted block, are denoted by $C_{k_1}(X_b)$. Under these assumptions, the kth intercepted symbol can be given as

$$r(k) = HS(k) + w(k), \quad (3)$$

where $S(k) = C_p(X_q)$, $p = (k+k_1) \bmod L$, $q = b + (k+k_1) \text{div} L$, and z mod L and z div L respectively denote the remainder and quotient of the division $z/L$. $w(k)$ represents complex Gaussian white noise with zero mean and variance $\sigma_w^2$. $H = [h_0, h_1, \cdots h_{N_t-1}]$ is the vector of fading channel coefficients, which is constant over the observation period.

In this paper, we apply a CNN to identify the STBC signal, which demonstrates that classification accuracy can be improved over current approaches.

*Difference analysis between SM signal and AL signal:* Equation (1) shows that the transmit SM signal in two consecutive time periods consists of IID random variables, whose correlation function is

$$E(x_0 x_1) = 0, \quad (4)$$

where $x_1$ and $x_2$ are the transmit signals in the first and second time period, respectively.

In the same way, for the AL signal, equation (2) shows a correlation between the two consecutive time periods. The correlation function is

$$E(x_0 x_1) = E(x_0 x_0^*) = x^2, \quad (5)$$

where ∗ denotes complex conjugation.

For an AL signal, the received signals in the two consecutive time periods are

$$r(0) = h_0 x_0 + h_1 x_1 + n_0 \quad (6)$$

$$r(1) = -h_0 x_0^* + h_1 x_1^* + n_1. \quad (7)$$

The correlation function is

$$E[r(0)r(1)] = h_1^2 x_1^2 - h_0^2 x_0^2. \quad (8)$$

For an SM signal, the received signals in the two consecutive time periods are

$$r(0) = h_0 x_0 + h_1 x_1 + n_0 \quad (9)$$

$$r(1) = h_0 x_3 + h_1 x_4 + n_1. \quad (10)$$

The correlation function is

$$E[r(0)r(1)] = 0. \quad (11)$$

From equations (4)-(11), we can find from the difference between the AL signal and SM signal that AL symbols are partly interrelated, while SM symbols are not. So, we apply the CNN to extract features in order to quickly identify the signal.

*Generating and Labeling the dataset:* A suitable dataset is necessary in subsequent experiments. Consequently, we select a collection of STBC signals that includes SM and AL signals, which are widely used in actual communication. We transmit a known signal that follows the coding matrices (1) and (2), and use MATLAB to expose them to the channel effects and noise effects described above. The samples are segmented by a short-time window, just as we segmented voice signals. The dataset is formed by extracting steps of 128 samples with a shift of 64 samples.

The labels that contain the information of SNR and the type of STBC are added in the first two columns. The dataset has 130 samples. The SNR is set between -10 and 10 dB, and the type of STBC is set as AL or SM.

*Preprocessing a dataset:* The CNN originally appears in image processing. An image that contains pixels is generally two-dimensional (2-D) data. A CNN generally deals with 2-D data, but the radio time series $r(t)$ is 1-D. Consequently, we must convert the 2-D dataset to 1-D. We use $X(t)$ as a new set of $2 \times N$ vectors, which extract in-phase (I) and quadrature (Q). Half of the sample is used for training and half for verification.

*Model building and training：* We select a four-layer network as a candidate after we train against several neural networks. The candidate neural network has several parts, consisting of two convolutional layers and two dense fully connected layers. The activation function is set as a rectified linear (ReLU) function in the first three layers; soft-max activation is used in the one-hot output layer. The CNN architecture is shown in Figure 1.

Regularization is used to prevent overfitting, and some neural network units are temporarily discarded from the network according to a certain probability. We use a dropout of 50% on the convolutional layer weights and the first dense layer. We run on a RTX2080ti GPU supported by NVIDIA CUDA in Keras on top of TensorFlow for network training and prediction.

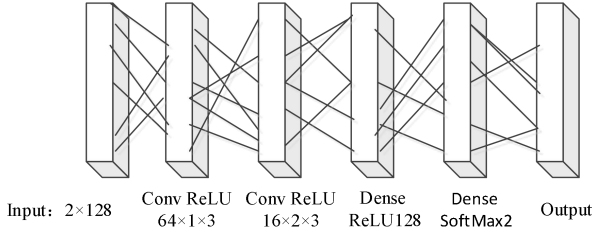

**Fig. 1** *CNN architecture*

After model building, the structure is as shown in Table I.

**Tabel 1:** Model Structure

| Layer (type) | Output shape | Parameter# |
| --- | --- | --- |
| Reshape_1 (Reshape) | (None,1,2 128) | 0 |
| Conv1 (Conv2D) | (None,256,2 129) | 1280 |
| conv2 (Conv2D) | (None, 80, 1, 131) | 122960 |
| dense1 (Dense) | (None, 256) | 2683136 |
| dense2 (Dense) | (None, 2) | 514 |

*Simulation setup:* To evaluate the proposed method, we verify the classification performance on the test dataset. The size of the dataset is on the order of millions. A Nakagami-m channel is adopted (m=3), and the noise is Gaussian white noise. QPSK modulation is used for simulation. The SNR ranges from -20 dB to +20 dB and is labeled to evaluate performance on each SNR condition.

*Performance evaluation:* Fig. 2 shows the probability of correct identification $P(\lambda = \xi | \xi), \xi = \text{AL,SM}$ versus SNR. The classification accuracy of the proposed algorithm reaches roughly 1 at SNR=-5 dB, hence it performs well even at a low SNR. To further illustrate the performance of the proposed algorithm, we compare it to existing expert feature-extraction algorithms. The performance of CNNs exceeds that of the expert feature extraction algorithms. The proposed algorithm requires no prior information of channel coefficients and noise power, and consequently is well-suited for a non-cooperative context. The validation loss and training loss are shown in Fig. 3. The curves begin to converge in 15 epochs, which means that the deep network works well and converges quickly.

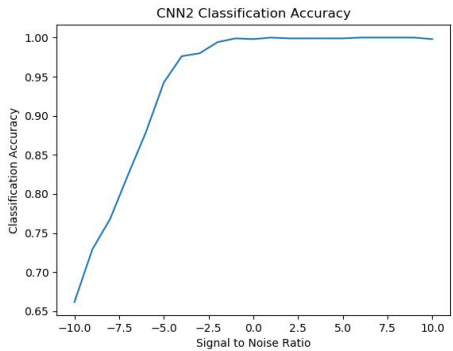

**Fig. 2** *Classifier performance vs SNR*

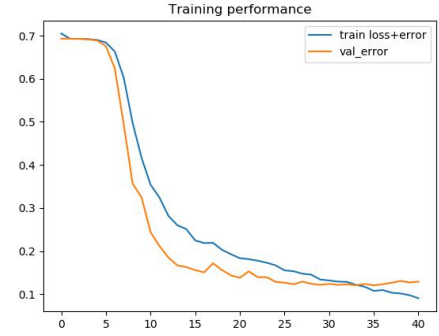

**Fig. 3** *Loss plots CNN2*

*Confusion matrix:* To further understand the performance of the proposed algorithm, we should seek its limit. We look at the confusion matrix at SNR=-5 dB, -10 dB, 0 dB, and 5 dB, as shown in Fig. 4-Fig. 7. When the SNR is greater than -5dB, the confusion matrix has a clean diagonal feature.

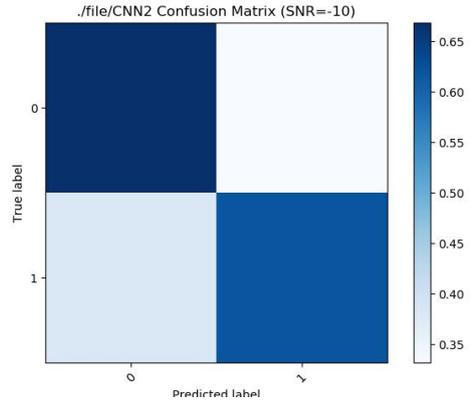

**Fig. 4** *-10 dB performance of CNN2*

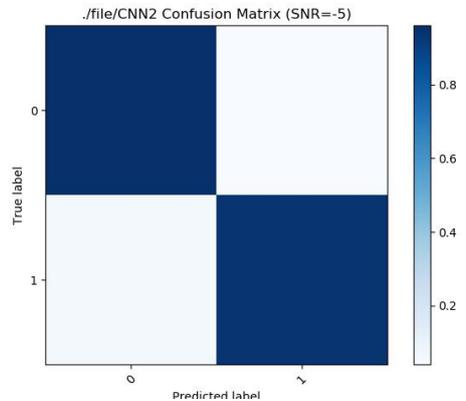

**Fig. 5** *-5 dB performance of CNN2*

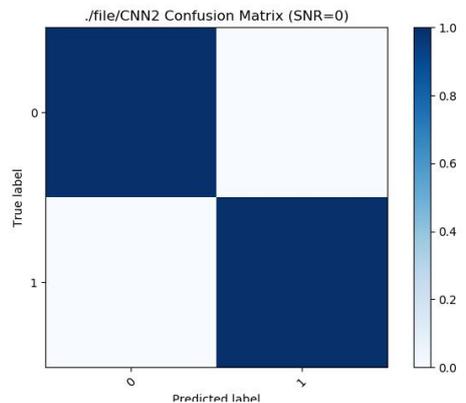

**Fig. 6** *0 dB performance of CNN2*

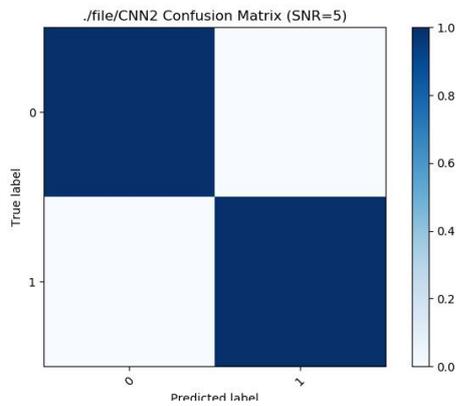

**Fig. 7** *5 dB performance of CNN2*

*Performance comparison:* We compare the performance of the proposed method with Q. Ling's algorithm[10], which uses higher-order cumulants (HOC) for classification, and the number of verification samples is 1,024. It can be seen from Fig. 8 that the recognition accuracy of this algorithm is close to 100% when the SNR is – 5 dB, while that of HOC is about 5 dB, and the recognition accuracy is close to 100%. The performance of HOC improves as the number of samples increases. The performance of this algorithm does not depend on the number of samples because its core is based on deep learning to build a network, and the recognition accuracy of the algorithm depends on the network, not on the number of verification samples. This is the advantage of deep learning.

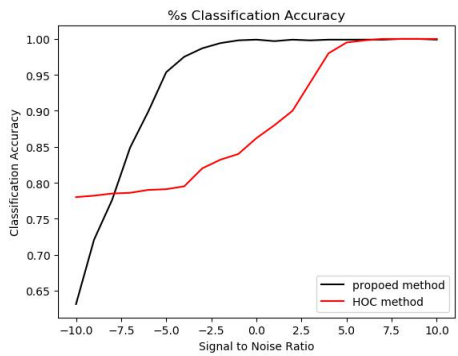

**Fig. 8** *Performance comparison*

*Conclusion:* We proposed a new algorithm to recognize AL and SM signals using CNNs. The algorithm provides reasonable performance at relatively low SNRs. Furthermore, it requires no prior information such as channel coefficient and noise power. Future work will include the extension of the deep neural network presented in this paper to other methods, such as residual networks, inception modules, and LSTM networks. We plan to extend the identification method to additional STBCs such as STBC-OFDM and SFBC-OFDM.

W. Yan, Q. Ling, L. Zhang and Y. Zhang (*Naval Aviation University, 264001 Yantai, China*)

E-mail: wj_yan@foxmall.com; linqing19870522@163.com